\documentclass{PoS}

\usepackage{amsmath}
\usepackage{amsfonts}
\usepackage{amssymb}
\usepackage{amsthm}
\usepackage{graphicx}
\usepackage{subcaption}
\usepackage{graphicx}
\usepackage{mathtools}
\usepackage{float}
\usepackage{bm}  
\usepackage{float}
\usepackage{relsize}
\usepackage{ulem}
\usepackage[shortcuts]{extdash}

\newcommand{\full}{\mathrm{full}}

\title{Open Effective Field Theories\\ from Highly Inelastic Reactions
\thanks{This work was supported in part by the Department of Energy, the Deutsche Forschungsgemeineschaft,
the National Science Foundation, and the Simons Foundation.}
}
\ShortTitle{EFT for Highly Inelastic Reactions}

\author{\speaker{Eric Braaten}
\\
        Department of Physics,
         The Ohio State University, Columbus, OH\ 43210, USA}
        
\author{Hans-Werner Hammer\\
Institut f\"ur Kernphysik, Technische Universit\"at Darmstadt,
64289\ Darmstadt, Germany}

\author{G.~Peter Lepage\\
Laboratory of Elementary Particle Physics, Cornell University,
Ithaca, New York\ 14583, USA}

\abstract{
The loss of particles due to highly inelastic reactions has
previously been taken into account in effective field theories for
low-energy particles by adding local anti-Hermitian terms to the effective
Hamiltonian.  An additional modification is required in 
the time evolution equation for the density matrix of a multi-particle system.  
An effective density matrix can be defined by tracing over states containing
high-momentum particles produced by the highly inelastic reactions
and by a time average that eliminates short-time correlations.
The effective density matrix satisfies the Lindblad equation, with local Lindblad operators
that are determined by the anti-Hermitian terms in the effective Hamiltonian. }

\FullConference{38th International Conference on High Energy Physics\\
		3-10 August 2016\\
		Chicago, USA}

\emergencystretch=\maxdimen
\begin{document}

\section{Highly Inelastic Reactions}

There are multiparticle systems consisting of elementary particles with reactions that are  {\it highly inelastic}: 
they produce particles with much larger momenta than the particles in the system \cite{BHL:1607}.
Some examples are
\begin{itemize}
\item
a muon beam, from which  muons can be lost by the decay  $\mu^- \to \nu_\mu \, e^- \, \bar \nu_e$,
\item
wimp dark matter, in which a pair of wimps can annihilate into electroweak gauge bosons,
\item
an axion star, in which low-energy axions can scatter into relativistic axions
through the reaction $a a a a \to a a$.
\end{itemize}
Many important loss processes for ultracold atoms also involve deeply inelastic reactions \cite{Braaten:2016dsw}.
Since the particles from the inelastic reactions have much larger momenta 
than the particles in the system, it should be possible to integrate them out 
and describe the particles of the system by an effective field theory.
What is the appropriate effective field theory?
The answer turns out to be somewhat surprising.
    
\section{Muon decay paradox}

For simplicity, we begin by considering a system consisting of nonrelativistic muons.
The muon decay products $\nu \, e \, \bar \nu$ have momenta of order $m_\mu$,
so the decay is highly inelastic.
The decay proceeds through a highly virtual $W$,
which  can propagate only over short distances of order $1/M_W$.
The $W$ can therefore be integrated out in favor of a contact interaction between $\mu$
and $\nu \, e \, \bar \nu$.
Thus the leptons can be described by a local effective field theory
with a 4-fermion interaction.
                         
The reaction $\mu \to \nu \, e \, \bar \nu$ is also local in another sense that may not be quite as familiar.                                                     
A high-momentum lepton created by the decay has a de Broglie wavelength of order  $1/m_\mu$,
so the creation of its wavepacket can be traced back to a region with size of order $1/m_\mu$.
The disappearance of the muon therefore occurs in a region with size of order $1/m_\mu$.
The effects of the reaction on muons can  be reproduced 
by a local operator that annihilates and recreates a muon at a point.
The effective field theory from integrating out the high-momentum leptons 
has a local non-Hermitian Hamiltonian:
\begin{equation}
H_\mathrm{eff}  =  H - i K.
\end{equation}
In a nonrelativistic effective field theory with muon field $\psi$, 
the Hermitian operator $K$ in the anti-Hermitian part is
\begin{equation}
K = \tfrac12 \Gamma_\mu \int d^3r\,  \psi^\dagger \psi  = \tfrac12 \Gamma N,
\end{equation}
where $N$ is the muon number operator.
Note that the muon number operator commutes with the effective Hamiltonian:
$[H,N] = 0$ and $[K,N] = 0$.

An effective density matrix $\rho$ for the system of low-energy muons can be defined
by tracing the density matrix $\rho_\mathrm{full}$ for the full system
over states that include high-momentum leptons from muon decays:
\begin{equation}
\rho  =  \mathrm{Tr}_\mathrm{high}(\rho_\mathrm{full}).
\label{rho-def}
\end{equation}
What is the time evolution equation for this density matrix? 
An obvious guess is the time evolution equation implied by the 
Schr\"odinger equation with the non-Hermitian Hamiltonian $H_\mathrm{eff}$:
\begin{equation}
i\hbar \frac{d\ }{dt}\rho \overset{?}{=}   
H_\mathrm{eff}\,  \rho - \rho\,  H_\mathrm{eff}^\dagger = 
\big[H, \rho \big]    - i \big\{K,\rho \big\}.
\label{evol-naive}
\end{equation}
This equation implies that the total probability $\mathrm{Tr}(\rho)$
is a decreasing function of time:
\begin{equation}
\frac{d\ }{d t} \mathrm{Tr}(\rho) \overset{?}{=}  - \Gamma_\mu  \mathrm{Tr}(N \rho) 
= - \Gamma_\mu \langle N \rangle.
\label{Tr(rho)-naive}
\end{equation}
The decreasing probability may not be disturbing, 
given that decays are steadily removing muons from the system.  
However the corresponding  equation  for the mean number  of muons,
$\langle N \rangle = \mathrm{Tr}(N \rho)$, is
\begin{equation}
\frac{d\ }{d t} \langle N \rangle \overset{?}{=}  - \Gamma_\mu  \mathrm{Tr}(N^2 \rho) 
= - \Gamma_\mu \langle N^2 \rangle.
\end{equation}
This implies that the muon number in a system with a large number $N$
of muons decreases at a rate $N$ times larger than the decay rate for a single muon.
This is in dramatic contradiction to our physical intuition that 
the number of muons in a dilute system must decrease exponentially with time
as $\exp(- \Gamma_\mu t)$.
Before presenting the resolution of this paradox,
we first make a detour into quantum information theory.

\section{Lindblad equation}

An arbitrary statistical ensemble of quantum states for a closed system
can be represented by a density matrix $\rho$.
The density matrix is Hermitian:  $\rho^\dagger = \rho$.
It is positive: $\langle \psi | \rho | \psi \rangle \ge 0$ for  all nonzero states $|\psi\rangle$.
These two properties ensure that if the density matrix is normalized so ${\rm Tr}(\rho) = 1$,
its eigenvalues can be interpreted as probabilities.  
If the system is in a pure quantum state, its density matrix satisfies  $\rho^2 = \rho$.
The system average of an operator $O$
can be expressed as the trace  of its product with the density matrix:
\begin{equation}
\langle O \rangle = {\rm Tr}(\rho O ).
\label{d<O>}
\end{equation}

The time evolution of the density matrix can be derived from
the Schr\"odinger equation for the quantum states:
\begin{equation}
i \hbar \frac{d\ }{dt} \rho =  [ H, \rho],
\label{evol:Schrodinger}
\end{equation}
where the Hermitian operator $H$ is  the Hamiltonian for the system.
This evolution equation has the following properties:
\begin{itemize}
\item[{\bf 1}.]
It is {\it linear}  in $\rho$.
\item[{\bf 2}.]
It {\it preserves the trace} of $\rho$.
The total probability ${\rm Tr}(\rho) = 1$ is therefore conserved.
\item[{\bf 3}.]
It is {\it Markovian}.    The future is determined by the present only, with
no additional dependence on the past history.
\end{itemize}
If the operator $O$ has no explicit time dependence,
the time evolution of the system average $\langle O \rangle$ is determined by
the evolution equation for $\rho$ in Eq.~\eqref{evol:Schrodinger}.

A quantum system that consists of a {\it subsystem} of interest and its {\it environment}
is often referred to as an {\it open quantum system}.
Of particular interest is the decoherence of the subsystem due to the effects of the environment.
An effective density matrix $\rho$ for the subsystem can be obtained 
from the density matrix $\rho_\full$ for the full system 
by the partial trace over the states of the environment.
The density matrix $\rho_\full(t)$ evolves in time according to  Eq.~\eqref{evol:Schrodinger}.
Given a specified initial state of the environment at time $t=0$, 
this evolution equation determines $ \rho(t)$ at future times.
It is possible in principle to construct a self-contained differential equation
for $ \rho(t)$, but it is nonlinear in $\rho$ and it is non-Markovian:
the time derivative of $\rho(t)$  is determined
by the present density matrix and by its history from time 0 to $t$.
The previous history is needed to take into account the flow of information
between the subsystem and the environment.

There are situations in which the time evolution of the effective density matrix $\rho$ for the subsystem
can be described approximately by a self-contained differential equation
satisfying the conditions {\bf 1}, {\bf 2}, and {\bf 3} itemized above.
The Markovian condition requires the time scale for an observation of the subsystem
to be larger than that for correlations between the subsystem and the environment.
This condition can be ensured by a suitable time average of the density matrix obtained 
by tracing over the environment.
In 1976, Lindblad and, independently, Gorini, Kossakowski, and Sudarshan showed that
the time evolution equation for $\rho$ is strongly constrained if the conditions  {\bf 1}, {\bf 2}, and {\bf 3}
are supplemented by one additional condition \cite{Lindblad:1975ef,Gorini:1975nb}:
\begin{itemize}
\item[{\bf 4}.]
The time evolution of $\rho$ is {\it completely positive}.
\end{itemize}
This condition requires that  if $\rho$ is extended to a density matrix 
on the tensor product of the subsystem and an arbitrary complex vector space,
that density matrix remains positive under time evolution  \cite{Preskill}.
Complete positivity ensures that the density matrix for any entangled state
of the subsystem and an environment with which it does not interact 
evolves into a density matrix that remains positive.
Given conditions {\bf 1}, {\bf 2}, {\bf 3}, and {\bf 4}, 
Lindblad and  Gorini, Kossakowski, and Sudarshan showed that the time evolution for $\rho$ 
must be given by the {\it Lindblad equation}  \cite{Lindblad:1975ef,Gorini:1975nb}:
\begin{equation}
i \hbar \frac{d\ }{dt} \rho =
 [ H, \rho] - \frac{i}{2} \sum_n
\left( L_n^\dagger L_n \rho + \rho L_n^\dagger L_n
-2 L_n \rho L_n^\dagger \right),
\label{Lindblad}
\end{equation}
where  $H$ is a Hermitian operator
and the $L_n$'s are an additional set of operators called {\it Lindblad operators}.
An accessible derivation of the Lindblad equation can be found in lecture notes 
by John Preskill on Quantum Computation  \cite{Preskill}.
The Lindblad equation can be expressed in the form
\begin{equation}
i \hbar \frac{d\ }{dt} \rho =
 [ H, \rho] - i \left\{K,\rho\right\}
 + i \sum_n L_n \rho L_n^\dagger ,
\label{Lindblad-2}
\end{equation}
where the Hermitian operator $K$ is
\begin{equation}
K =\frac12\sum_n L_n^\dagger L_n.
\label{eq:K-Ln}
\end{equation}
Comparison with the naive evolution equation in Eq.~\eqref{evol-naive}
reveals that $H - i K$ can be interpreted as a non-Hermitian effective Hamiltonian 
for the subsystem. The additional {\it Lindblad term} in Eq.~\eqref{Lindblad-2},
whose form is determined by that of $K$, ensures that Tr($\rho$) is conserved.

\section{Muon decay revisited}

We now present the solution to the muon decay paradox presented above.
An effective density matrix can be defined by the partial trace in Eq.~\eqref{rho-def}
It can be made at least approximately Markovian by a time average
that eliminates transients associated with muon decays,
provided the muon decay products subsequently escape from the system.
If we impose conditions {\bf 1}, {\bf 2}, {\bf 3}, and {\bf 4} as reasonable physical conditions 
on the  effective density matrix,
its time evolution must be given by the Lindblad equation:
\begin{equation}
\frac{d\ }{d t} \rho = -i [ H, \rho] 
- \tfrac{1}{2} \Gamma_\mu  \int d^3r \,\big( \psi  \psi^\dagger\,  \rho + \rho\, \psi  \psi^\dagger 
- 2 \psi  \rho  \psi^\dagger \big).
\label{evol-Lindblad}
\end{equation}
The trace of the right side is 0, so the total probability Tr($\rho$) is conserved.
Using the commutation relations $[N,\psi]= -\psi$ and $[N,\psi^\dagger]= +\psi^\dagger$,
the time derivative of $\langle N \rangle = \mathrm{Tr}(N \rho)$  can be reduced to
\begin{equation}
\frac{d\ }{d t} \langle N \rangle =  - \Gamma_\mu  \mathrm{Tr}(N \rho) 
= - \Gamma_\mu \langle N \rangle.
\end{equation}
Thus $\langle N \rangle$ decreases like $\exp(-\Gamma_\mu t)$,
 in accord with our physical intuition.

Some insights into the Lindblad equation can be obtained
by considering the probability $P_n(t)$ for $n$ muons,
which can be expressed as a partial trace of $\rho$:  
$P_n = \mathrm{Tr}_{N=n} (\rho)$.
Taking the partial trace of the Lindblad equation, we obtain the evolution equation
\begin{equation}
\frac{d\ }{d t} P_n =
-\Gamma_\mu  \big[ n P_n - (n+1) P_{n+1} \big].
\label{evol-Pn}
\end{equation}
The term proportional to $P_{n+1}$ comes from the Lindblad term in the Lindblad equation.
If that term is omitted, Eq.~\eqref{evol-Pn} for $P_n$ is essentially that for Tr($\rho)$ in Eq.~\eqref{Tr(rho)-naive}.
Thus the naive evolution equation in Eq.~\eqref{evol-naive} can be interpreted as that for a density matrix 
obtained by projecting onto states that include no high-momentum particles from inelastic reactions.
That density matrix has a probability that decreases rapidly  with time, 
and it quickly becomes physically irrelevant.
The physically relevant density matrix is obtained instead by tracing over states that include 
high-momentum particles, and it satisfies the Lindblad equation in Eq.~\eqref{evol-Lindblad}.

\section{Open effective field theory}

It is straightforward to generalize our result for muons 
to any system with highly inelastic reactions.
The effective Hamiltonian from integrating out the high-momentum
particles from the inelastic reactions has the form
$H - i K$, where the Hermitian operator $K$ is local: 
\begin{equation}
K = \sum_i \gamma_i \int d^3r\,  \Phi_i^\dagger \Phi_i.
\end{equation}
The local operator $\Phi_i$ annihilates low-energy particles 
in a configuration that corresponds to the initial state of a highly inelastic reaction.
The operator $K$ is positive if the coefficients $\gamma_i$ are positive.
An effective density matrix $\rho$ can be defined by tracing over states 
that include high-momentum particles from the highly inelastic reactions
and then averaging over a time long enough to eliminate transients associated with the decays.
The time evolution of the effective density matrix is described by the Lindblad equation:                     
\begin{equation}
\frac{d\ }{d t} \rho =
-i [ H, \rho] -  \left\{K,\rho\right\}
 + 2 \sum_i \gamma_i \int d^3r \, \Phi  \rho  \Phi^\dagger .
\end{equation}

Burgess, Holman, Tasinato, and Williams have shown that the effective density matrix 
 for super-Hubble modes of primordial quantum fluctuations in the early universe 
 satisfies a Lindblad equation  \cite{Burgess:2014eoa}.
 They proposed the terminology {\it open EFT} for an effective field theory 
in which the density matrix satisfies the Lindblad equation.
Thus the effective field theory from integrating out high-momentum particles 
from highly inelastic reactions is an open EFT.  The low-energy particles
are the subsystem of interest in the open quantum system, 
and the high-momentum particles are its environment. 

The concluding sentence of Ref.~\cite{Burgess:2014eoa}  was
``Now that we have a new hammer,  let's go find all those nails \ldots''.
One such nail is the loss of low-energy axions from an axion star, 
which is a gravitationally bound Bose-Einstein condensate of axions.
In Ref.~\cite{Eby:2015hyx}, the authors proposed a new loss mechanism in which 
3 condensed axions make a transition to a single relativistic axion through the 4-axion vertex.
The condensed axions can be described by a nonrelativistic effective field theory 
with a complex field $\psi$.
The 4-axion vertex allows the scattering of 3 low-energy axions through a single virtual axion,
but the amplitude has no imaginary part.  Thus there is no anti-Hermitian term 
 of the form $(\psi^3)^\dagger \psi^3$ in the effective Hamiltonian density for the open EFT.
Instead the leading mechanism for producing relativistic axions is the scattering 
of 4 condensed axions to 2 relativistic axions \cite{Braaten:2016dlp}.
The one-loop diagrams  for the scattering of 4 low-energy axions have imaginary parts
that correspond to a local anti-Hermitian term  in $H_\mathrm{eff}$
of the form $(\psi^4)^\dagger \psi^4$.


\end{document}